\documentclass[useAMS,usenatbib]{mn2e}
\usepackage{amsmath,amssymb}
\usepackage[a4paper,bindingoffset=0.2in,%
            left=0.3in,right=0.5in,top=1in,bottom=1in,%
            footskip=.25in]{geometry}
\usepackage{graphicx}
\usepackage{color}

\newcommand{\beq}{\begin{equation}}
\newcommand{\eeq}{\end{equation}}

\newcommand{\dd}{\partial}

\title{LISA Detection of Binary Black Holes in the Milky Way Galaxy}

\author[Pierre Christian and Abraham Loeb]{Pierre Christian and Abraham Loeb \\
Harvard Smithsonian Center for Astrophysics, 60 Garden Street, Cambridge, MA, U.S.A.
}

\begin{document}
\maketitle
\begin{abstract}
Using the black hole merger rate inferred from LIGO, we calculate the abundance of tightly bound binary black holes in the Milky Way galaxy. Binaries with a small semimajor axis ($\lesssim 10 R_\odot$) originate at larger separations through conventional formation mechanisms and evolve as a result of gravitational wave emission. We find that LISA could detect 
them in the Milky Way. We also identify possible X-ray signatures of such binaries.  
\\
\end{abstract}


\section{Introduction}
The Laser Interferometer Gravitation-Wave Observatory (LIGO) discovered gravitational waves from binary black holes \citep{LIGO1, LIGO3}, composed of black holes with masses $\gtrsim 10 M_\odot$. 

Two stars in an isolated binary can evolve to produce the progenitors of the LIGO sources. To possess the observed parameters of the LIGO sources, these binaries must have progenitors with high masses ($M \sim 40 - 100 M_\odot$) and low metallicities \citep{Belczynski}. The binary evolution could be initially affected by mass transfer through a common envelope phase.

However, in the chemically homogeneous evolution model \citep{Mandel1, Mandel2}, two massive stars in a near contact binary spin rapidly due to tidal spin-orbit coupling. The rapid rotation of a star mixes its interior, allowing transport of hydrogen from the envelope to the core and metals from the core to the envelope \citep{Maeder}. In contrast to the standard binary evolution model, the stars do not follow a common envelope phase, due to their contraction within their Roche lobes.

Regardless of the formation mechanism, none of the progenitor systems could directly produce two black holes at arbitrarily small separations. Here we focus on tight binary black holes with a semimajor axis smaller than could feasibly be created directly by conventional stellar evolution mechanisms. These binary black holes were born at a larger semimajor axis through standard  evolution, and then migrated to smaller separations via gravitational wave emission. Most of these binary black holes reside in an intermediate regime, where their coalescence time is shorter than the Hubble time but longer than the LIGO operation lifetime before they become detectable in the LIGO frequency band.

We focus our analysis on binary black holes within the Milky Way galaxy. While most of our equations could be applied to binaries at arbitrary distances, the signatures of the systems under consideration are not observable outside of the Milky Way. We will assume circular orbits, as the detected LIGO binaries are constrained to possess low eccentricities \citep{LIGO1}. This means that we neglect binary black hole production via many body encounters and strong interactions in globular clusters \citep{Lars, Rodriguez}. Aside from these assumptions, we will remain agnostic as to the specific mechanism producing the binaries.

Recently, \cite{Seto} used an estimate of the population of Galactic binary black holes to predict that LISA will have the sensitivities required to detect binaries like GW150914. In this article, we used a more sophisticated population analysis to show that the result can be generalized to more complicated population models. Furthermore, we extend the calculation to black hole populations with varying masses and take into account the black hole mass function. 

In section 2 we perform a population analysis of tight binary black holes in the Milky Way. In section 3 we show that the gravitational wave signatures of these tight binaries are observable by LISA. In section 4 we examine possible X-ray signatures of these binaries. Finally, section 5 summarizes our conclusions. 

\section{Population analysis}

The number of binary black holes at a given time $t$ with semimajor axis between $a$ and $a+da$ can be written as,
\beq
dN(a, t) = \rho(a, t) da \; . 
\eeq
Assuming that their dynamical evolution is dominated by the emission of gravitational radiation, $\rho(a,t)$ obeys  a simple advection equation, 
\beq \label{eq:advec}
\frac{\dd \rho(a,t)}{\dd t} - \frac{\dd }{\dd a} \left[ K(M_1, M_2) \frac{\rho (a,t)}{a^3} \right] = S(a, t) \; ,
\eeq
where $M_1$ and $M_2$ are the masses of the two black holes, $S(a,t)$ is a source term that parameterizes the production of binaries at a semimajor axis $a$, and $K$ is given by,
$$
K (M_1, M_2) \equiv \frac{64}{5} \frac{G^3}{c^5} (M_1 M_2) (M_1 + M_2) \; .
$$
In the simple case of $S=0$, the solution of equation (\ref{eq:advec}) is given by 
\beq \label{eq:homSol}
\rho(a,t) = a^3 F\left[ \frac{a^4}{4 K} + t \right] \; ,
\eeq
where $F$ is some arbitrary function. The simplest  solution can be obtained in a steady state, where equation (\ref{eq:advec}) reduces to 
\beq
\frac{\dd }{\dd a} \left[ K \frac{\rho(a)}{a^3} \right] = 0  \;. 
\eeq
Integrating this equation gives 
\beq
\rho(a) = C \frac{a^3}{K}  \; ,
\eeq
where $C$ is an arbitrary constant. This is a special case of the solution in equation (\ref{eq:homSol}), with $F=C/K$. 

The inferred merger rate from LIGO for two $30 M_\odot$ black holes is between $2$ and $600$ Gpc$^{-3}$ yr$^{-1}$ \citep{LIGO2} in comoving units. We can estimate the Galactic merger rate by adopting the number of Milky Way-like galaxies to be $10^{-2}$ per comoving Mpc$^{3}$ \citep{MWEG2}. Adopting $100$ Gpc$^{-3}$ yr$^{-1}$ as a fiducial LIGO inferred merger rate gives the LIGO Galactic merger rate to be $\sim 10^{-5} R_{100}$ mergers per galaxy per year. 

A merger is detected by LIGO when the semimajor axis of the binary is small enough that it enters the LIGO frequency band. Denoting this critical semimajor axis as $a_m$, the merger rate $R$ is equal to the flux in $a$-space at $a=a_m$,
\beq \label{eq:rhoMerg}
\rho(a_m) \frac{K}{a_m^3} = R \; .
\eeq
Therefore, 
\beq
\rho(a) = \rho(a_m) \left[ \frac{a}{a_m} \right]^3 = \frac{R}{K} a^3 \; ,
\eeq
where we have substituted $\rho(a_m)$ from equation (\ref{eq:rhoMerg}). Using the inferred Galactic merger rate and specializing to binary black holes of mass $M_1 = M_2 = 30 M_{\odot}$, we can integrate over $a$ to obtain the number of Galactic binary black holes with semimajor axis $\le a$,
\beq \label{eq:numberBHsimplest} 
N(\le a) \approx 3 \times 10^{-2} R_{100} \left[\frac{a}{\rm{R_\odot}} \right]^4 \; ,
\eeq
where $R_\odot$ is the solar radius and $R_{100}$ is the rate in units of $100\; \rm Gpc^{-3} \; yr^{-1}$.
  
\subsection{Source functions with a minimum injection scale} \label{subsection:SF with min scale} \label{subsection:maxinject}
Next, we generalize the population analysis to cases with a source function. Most formation mechanisms cannot produce binaries that are very tight. We model this situation with a source function that is proportional to a step function, $S = \tilde{S}(a) \Theta(a-a_0)$, where $a_0$ is the minimum binary separation for the formation mechanism, $\Theta$ is the Heavyside step function, and $\tilde{S}(a)$ is an arbitrary function. 

Given this source function, the general solution is, 
\beq \label{eq:solMinInject}
\rho(a,t) = -\frac{1}{K} a^3\left[ A \Theta(a- a_0)  \int_{a_0}^{a} \tilde{S} da      \right] + a^3 F\left[ \frac{a^4}{4 K} + t \right] \; ,
\eeq
as long as the function $\tilde{S}(a)$ is not singular at $a=a_0$. For example, a power law source with normalization $A$ and index $n$, $S = A a^n \Theta(a-a_0)$, admits the general solution,
\beq 
\rho(a,t) = -\frac{1}{K} a^3\left[ A  \frac{\Theta(a- a_0)}{(n+1)}  (a^{n+1} - a_0^{n+1})   \right] + a^3 F\left[ \frac{a^4}{4 K} + t \right] \; . 
\eeq
Note also that in the case of a delta function source injected at $a=a_0$, the solution is given by,
\beq
\rho(a,t) = -\frac{1}{K} a^3 A \Theta(a- a_0) + a^3 F\left[ \frac{a^4}{4 K} + t \right] \; . 
\eeq
The most important feature of equation (\ref{eq:solMinInject}) is that the solution below the injection point, $a < a_0$ is unchanged from the sourceless case. This implies that as long as we restrict our analysis to $a \le a_0$, we can simply use the sourceless solution. Our results will therefore be robust due to its insensitivity to the particular binary black hole production mechanism. 
  
\subsection{Power-law source functions with a maximum injection scale}
For the sake of generality, we also consider a source function without a minimum scale, namely $S(a) \propto a^n$ with positive $n$ that extends all the way to $a=0$. In principle, $n$ is related to the power law index of the binary separation of massive stars \citep{sana}. However, as not all massive binaries evolve into binary black holes, the mapping between the two indices is unknown. Since binaries are not produced up to arbitrarily high semimajor axis, we truncate our source function at high values of $a$ by an exponential factor, 
\beq
S(a) \propto a^n \exp{\left[-\frac{a}{a_c}\right]} \; ,
\eeq
where $a_c$, is some large semimajor axis above which binary black holes are rarely produced. This  source function corresponds to a mechanism that produces binaries over a broad range of $a$, where instead of a minimum injection scale, $a_0$, we now have a maximum injection scale, $a_c$. At large semimajor axes, this distribution corresponds to the end states of binary star evolution, whereas at small $a$ it corresponds to more exotic processes such as direct collapse \citep{Loeb2}. 

The binary population with this source term obeys
\beq \label{eq:advecsource}
\frac{\dd \rho(a,t)}{\dd t} - \frac{\dd }{\dd a} \left[ K(M_1, M_2) \frac{\rho (a,t)}{a^3} \right] = K_2 a^n \exp{\left[-\frac{a}{a_c}\right]} \; ,
\eeq
where $K_2$ and $n$ are constants that in principe can be constrained by observations. The general solution to this equation is given by,
\begin{equation}
\rho(a,t) = a^3 F\left[\frac{a^4}{4 K}+t\right] + \frac{a^{3} a_c^{n+1} K_2}{K} \Gamma\left[1+n, \frac{a}{a_c}\right]  \; ,
\end{equation}
where $F$ is an arbitrary function and $\Gamma$ is the incomplete Gamma function. As before, we can look for a steady state solution by setting the function $F$ to be a constant $C$, giving 
\beq
\rho(a) = a^3 C + \frac{a^{3} a_c^{n+1} K_2}{K}\Gamma\left[1+n, \frac{a}{a_c}\right]  \; .
\eeq
In this case, the merger rate $R$ equals the flux in $a$-space at $a=a_m$ plus a term corresponding to the source,
\beq
\rho(a_m) \frac{K}{a_m^3} + \sigma = R \; ,
\eeq
where $\sigma$ is the rate of binary black holes created with $a \le a_m$, given by,
\begin{align}
\sigma &= \int_0^{a_m}  K_2 a^n \exp{\left[-\frac{a}{a_c}\right]}  da \nonumber
 \\& = K_2 a_c^{n+1} \left[ \Gamma(1+n) - \Gamma\left(1+n, \frac{a_m}{a_c}\right) \right] \; .
\end{align}
Since $a_m \ll a_c$,  we get,
\beq
C = \frac{R - K_2 a_c^{n+1} \Gamma \left(1+n, a_m/a_c \right) }{K} \; .
\eeq
The number density of binary black holes is then given by,
\begin{align}
\rho(a) &= a^3 \frac{R}{K} + \frac{a^3 a_c^{n+1} K_2}{K}  \left[ \Gamma\left(1+n, \frac{a}{a_c}\right) -  \Gamma\left(1+n, \frac{a_m}{a_c}\right) \right] \nonumber
\\ &\approx a^3  \frac{R}{K} \; ,
\end{align}
where in the second equality we used the fact that both $a/a_c$ and $a_m/a_c$ are much smaller than unity. This result shows that in the case of a power law source function, the scaling of the sourceless solution $\rho \propto a^3$ remains valid even if the source function does not have a minimum scale as long as there is a maximum injection scale. The case of a power law with neither a maximum or minimum injection scale is treated in the Appendix. 

\subsection{Population analysis for varying black hole masses}
In the previous sections, the abundance of binaries was derived for a given value of $K(M_1, M_2)$. In reality, the black hole population spans a range of masses with a probability given by the black hole mass function. Thus, the binaries possess varying values of $K$. In this section we incorporate this diversity of $K$ values. Note that there is a single advection equation for every value of $K$, i.e. there are many copies of equation (\ref{eq:advec}), each for a different value of $K$. To make explicit its dependence on $K$, we label the density $\rho$ in equation (\ref{eq:advec}) as $\rho_K(a, K)$. The total number of binary black holes per semimajor axis is given in terms of $\rho_K(a,K)$ by,
\beq
\rho(a) = \int_{K_{min}}^{\infty} f_K \rho_K(a, K) dK  \; ,
\eeq
where $f_K$ is the probability of finding a binary black hole with a particular value of $K$. As $K=K(M_1, M_2)$ is a function of the two black hole masses, its probability distribution is determined by the mass functions of the first and second black holes, $f_{M_1}$ and $f_{M_2}$, respectively. For simplicity, we adopt $f_{M_1} = f_{M_2} = \Phi_M$, and the distribution $f_K$ can be obtained from $\Phi_M$ by a series of convolutions.
Adopting a phenomenological power-law relation, $f_K = K_3 K^m$ with an index $m < -1$ and a normalization,
\beq
\int_{K_{min}}^\infty f_K dK = 1 \; ,
\eeq
we can derive the normalization constant $K_3$ in terms of $K_{min}$. Theoretically, the minimum $K$ value is obtained when both black holes are at the limit imposed by the Tolman-Oppenheimer-Volkoff equation of a Chandrasekhar-Landau mass ($\sim 3 M_{\odot}$) each. However, there is evidence that there exists a mass gap under $\sim 5 M_\odot$ \citep{5M1,5M2,5M3}. We thereby chose the minimum $K$ to be that when both black holes are $\sim 5 M_\odot$ each.

For our phenomenological model with $m< -1$, the integral converges and can be solved to give,
\beq
K_3= \frac{ |m+1|}{K_{min}^{m+1}} =  |m+1| K_{min}^{|m+1|} \; .
\eeq
The merger rate is given by the sum over all masses of the fluxes in $a$-space,
\beq
\int_{K_{min}}^{\infty} f_K \rho_K(a_m, K) \frac{K}{a_m^3}   dK = R \; .
\eeq 
For the sourceless steady-state solution, $\rho(a) = Ca^3/K$, we can find $C$ in the phenomenological $f_K = K_3 K^m$ model by noting that,
\begin{align}
R & = \int_{K_{min}}^{\infty} K_3 K^m \left[ \frac{C a_m^3}{K} \right] \frac{K}{a_m^3}   dK \nonumber
\\ &= \frac{C K_3}{(m+1)} \left. K^{m+1}  \right\vert_{K_{min}}^{\infty} \; .
\end{align}
For $(m+1) < 0$, this integral converges to 
\beq
R = -\frac{CK_3}{(m+1)} K_{min}^{m+1} \; ,
\eeq
which implies that for $(m+1) < 0$,  
\beq
\rho_K(a, K) = -\frac{R  (m+1)}{ K_3 K_{min}^{m+1} } \frac{a^3}{K} \; .
\eeq
The number of binary black holes per unit semimajor axis is therefore given by,
\begin{align}
\rho(a) &=  -\frac{R  (m+1) a^3}{ K_3K_{min}^{m+1}} \int_{K_{min}}^{\infty} \frac{K_3 K^m}{K} dK \nonumber
\\ & =  \frac{(m+1)}{m} \frac{R a^3}{K_{min}}\; .
\end{align}
The number of Galactic binary black holes with semimajor axis $\le a$ is given by 
\begin{align}
N(\le a) &= \frac{(m+1)}{m} \frac{R a^4}{4 K_{min}} \; .
\end{align}
 
Aside from numerical factors of order unity the only change from the single $K$ case is that $K_{min}$ appears in the denominator in place of $K$. Since $K_{min}$ is $\sim 200$ times smaller than the $K$ for two $30$ solar mass black holes, this number is of order $200$ larger than in equation (\ref{eq:numberBHsimplest}).

\subsection{Population analysis for varying black hole masses: Chabrier/Kroupa IMF}

Next, we proceed beyond the phenomenological toy model for $f_K$ assuming that the black hole mass function follows the power-law dependence of massive stars, $\Phi_M = k M^{-2.3}$ \citep{Chabrier}, where $k$ is a constant. We define the quantity, 
\beq
\tilde{K} \equiv \frac{5 c^5}{64 G^3} K = (M_1 + M_2) M_1 M_2\; , 
\eeq
so that the respective distribution $f_{\tilde{K}}$ follows,
\beq
f_K (K) =   \frac{5 c^5}{64 G^3}  f_{\tilde{K}} \left[\frac{5 c^5}{64 G^3} K \right] \; .
\eeq
If the masses $M_1$ and $M_2$ are independently distributed, $f_K$ can be derived from $\Phi_M$ as follows. We first switch from the random variables $M_1$ and $M_2$ to $\tilde{K}$ and $W$, where $W \equiv M_1$ and,
\beq
M_2 = \frac{-W^2 + \sqrt{W^4 + 4 W \tilde{K} }}{2 W} \; ,
\eeq
where the positive root was chosen since $W$, $M_2$, and $\tilde{K}$ are positive definite. The distribution function $f_{\tilde{K} W}$ is then given by,
\beq
f_{\tilde{K} W} = |J| \Phi_M(W) \times \Phi_M\left[ \frac{-W^2 + \sqrt{W^4 + 4 W \tilde{K} }}{2 W} \right]  \;,
\eeq
where the determinant of the Jacobian of the transformation,
\beq
|J| = \frac{1}{\sqrt{W^4+4 W \tilde{K} } } \; .
\eeq
The marginal distribution $f_{\tilde{K}}$ is therefore,
\beq
f_{\tilde{K}} = \int_{M_{\rm min}}^{M_{\rm max}} f_{\tilde{K} W} dW \; , 
\eeq
which for the assumed mass function is given by,
\beq \label{eq:fKChab}
f_{\tilde{K}} = \int_{W_{\rm min}}^{W_{\rm max}} k^2 |J| W^{-2.3} \left[ \frac{-W^2 + \sqrt{W^4 + 4 W \tilde{K} }}{2 W} \right]^{-2.3} dW \; .
\eeq
Here, $W_{\rm min/max}$ corresponds to the minimum and maximum black hole masses; in particular, $W_{\rm min}$ is again chosen to be $\sim 5 M_\odot$ and $W_{\rm max} \sim 100 M_\odot$, respectively.

\begin{figure}
  \centerline{
   \includegraphics[scale=1]{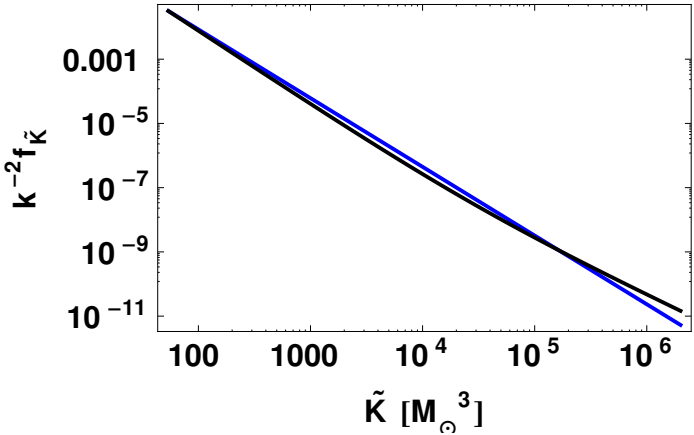} }
  \caption{\label{fig:fKChab} The probability function $f_{\tilde{K}}$ for the Chabrier/Kroupa IMF, $\Phi_M (m)= k m^{-2.3}$ (black), compared with the fitting function $A \tilde{K}^{-\alpha}$ for $A = 159$ and $\alpha=2.14$ (blue).} 
\end{figure} 

Although the integral in equation (\ref{eq:fKChab}) could only be solved numerically, the distribution is well represented by a power law form between $W_{\rm min}$ and $W_{\rm max}$ with a negative index $-\alpha \sim -2.1$ (see Figure \ref{fig:fKChab}). In order to simplify the analysis, we will therefore adopt, 
\beq
f_{\tilde {K}} (\tilde{K}) \approx k^2 A \tilde{K}^{-\alpha} \; ,
\eeq
where $A = 159$. This gives
\begin{align} 
f_K(K) &\approx \left[\frac{5 c^5}{64 G^3} \right]^{2/3} k^2 A K^{-\alpha} 
\\ &\equiv K_3 K^{-\alpha} \; .
\end{align}
We proceed analogously to the previous section, where the main difference is that we now have an upper cutoff on black hole masses at $W_{\rm max}$. 

For the steady state solution, $\rho(a) = Ca^3/K$, the rate of binary black holes entering the LIGO band is given by,
\begin{align} \label{eq:RChab}
R &= \int_{K_{min}}^{K_{max}} f_K \rho_K(a_m, K) \frac{K}{a_m^3}   dK \nonumber 
\\ &=  \int_{K_{min}}^{K_{max}} C K_3 K^{-\alpha} dK  \nonumber
\\  &= C K_3 \frac{(K_{max}^{1- \alpha} - K_{min}^{1 - \alpha} ) }{1- \alpha} \; .
\end{align}
In equation (\ref{eq:RChab}), the integration limits refer to the maximum and minimum black hole masses that LIGO is sensitive to. The assumption that LIGO is sensitive to all black hole masses available in the black hole mass function translates to substituting $K_{max}$ and $K_{min}$ as these limits. Note that as the IMF is dominated by low mass black holes, our result is only weakly dependent on the exact value of $K_{max}$. Using $R$ to eliminate the constant $C$ yields,
\beq
\rho_K (a, K) =  \frac{R (1 - \alpha) }{ K_3(K_{max}^{1 - \alpha} - K_{min}^{1 - \alpha} ) } \frac{a^3}{K} \; .
\eeq
The abundance of binaries is therefore given by,
\begin{align}
\rho(a) &=  \int_{K_{min}}^{K_{max}} f_K \rho_K(a, K) dK \nonumber
\\ &=  \frac{(1- \alpha) R a^3}{(K_{max}^{1 - \alpha} - K_{min}^{1 - \alpha} ) } \int_{K_{min}}^{K_{max}} \frac{K^{- \alpha}}{K} dK  \nonumber
\\ &=  \frac{(1 - \alpha) R a^3 (K_{min}^{-\alpha} - K_{max}^{-\alpha})  }{ \alpha (K_{max}^{1- \alpha} - K_{min}^{1 - \alpha} ) } \; .
\end{align}
The number of binary black holes with semimajor axis $\le a$ is then,
\begin{align} \label{eq:NumIMF}
N(\le a) &= \frac{(1 - \alpha) R a^4 (K_{min}^{-\alpha} - K_{max}^{-\alpha})  }{ 4 \alpha (K_{max}^{1- \alpha} - K_{min}^{1 - \alpha} ) }
\\ &\approx 3 \times 10^{4} R_{100} \left[\frac{a}{10 R_\odot} \right]^4 \; . 
\end{align}
While this result is derived by assuming that the black hole mass function follows the power-law mass function of massive stars, simulations indicate that not all massive stars form black holes, and that stars above $\sim 50 M_\odot$ blow off too much of their mass to produce black holes \citep{Sukhbold}. These complications could lower the abundance of black holes relative to that predicted by equation (\ref{eq:NumIMF}). This means that our prediction should be interpreted as an upper bound to the number of binary black holes in the Galaxy.

\subsection{Comparison with population synthesis models}
In order to compare our result to population synthesis models, we first transform our variable from $a$ to the frequency $f$. The number of binary black holes with semimajor axis $\le a_{\rm max}$ is equal to the number of binary black holes with frequency $\ge f_{\rm min}$, where $f_{\rm min}$ is the orbital period of the smallest black holes in the population with separation $a_{\rm max}$. 

For the systems under consideration, $a_{\rm max} = 10 R_{\odot}$ and the smallest black hole mass is $5 M_\odot$, resulting in a minimum frequency of $f_{\rm min} = 2 \times 10^{-5}$Hz. Equation (\ref{eq:NumIMF}) therefore predicts $\approx 3 \times 10^{4} R_{100} $ binary black holes in the Galaxy with frequency greater than $2 \times 10^{-5}$Hz.

The population synthesis model A of \cite{Benacquista} predicts $\sim$thousands of binary black holes with frequencies greater than $2 \times 10^{-5}$Hz. Noting that the LIGO rate $R_{100}$ ranges from $0.02$ to $6$, this is consistent with our result. An earlier calculation by \cite{Nelemans} predicts a number that is an order of magnitude larger than model A of \cite{Benacquista}, which is still consistent with our result.

\section{Gravitational wave signal from Milky Way binaries}

Most of the $3\times 10^4 R_{100}$ Galactic binary black holes with orbital separation $a \le 10 R_\odot$, will not enter the LIGO bandpass in a short enough time for them to be observed by LIGO. For example, the timescale for two $30 M_\odot$ black holes to coalesce from $a \sim$ a few $R_\odot$ is thousands of years. These  binaries, however, will be observable by LISA\footnote{http://www.elisascience.org} which is sensitive to lower frequencies than LIGO.

\begin{figure}
  \centerline{
   \includegraphics[scale=0.6]{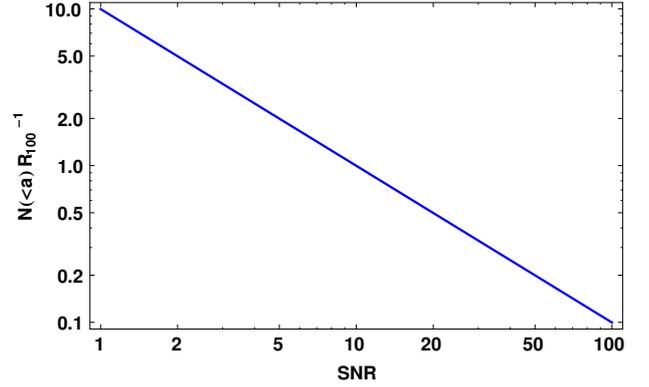} }
  \caption{\label{fig:LISA} Expected number of Milky Way binaries composed of two $30 M_\odot$ black holes as a function of the SNR at a distance $d = 20$ kpc.} 
\end{figure}

Focusing on the case of two $\sim 30 M_\odot$ black holes, we find from equation (\ref{eq:numberBHsimplest}) that the tightest binary black hole in our Galaxy has $a \sim 2.5 R_\odot$. For such a binary consisting of two $30 M_\odot$ black holes, the gravitational wave frequency is $f \sim 3 \times 10^{-4}$ Hz, which is within the LISA bandpass \citep{WDbackground}. The angular-averaged gravitational wave strain for the $n=2$ mode is given by \citep{Peters1, Seto},
\beq
A \approx 2.1 \times 10^{-20} \left( \frac{8 \; \rm kpc}{d} \right) \left( \frac{M_c}{28 M_\odot} \right)^{5/3} \left( \frac{f}{5 \times 10^{-4} \; \rm Hz 	} \right)^{2/3} \; ,
\eeq
where $M_c \equiv (M_1 M_2)^{3/5}(M_1 + M_2)^{-1/5}$ is the chirp mass. Integrating the signal over an observational period $\tau$, the signal to noise ratio becomes \citep{Seto},
\beq
SNR \approx 70 \left( \frac{A}{2.1 \times 10^{-20}} \right) \left( \frac{h(f)}{3 \times 10^{-18} \; \rm Hz^{-1/2}} \right)^{-1} \left( \frac{\tau}{3 \; \rm years} \right)^{1/2} \; , 
\eeq
where $h(f)$ is the LISA instrumental noise, with a value of $\sim 3 \times 10^{-18} \; \rm{Hz^{-1/2}}$ at $0.5$ mHz \citep{Amaro}. Scaling the noise with frequency as the power law $h(f) \propto f^{-2}$ \citep{Seto}, the signal to noise ratio becomes,
\begin{align}
SNR &\approx 70 \left( \frac{A}{2.1 \times 10^{-20}} \right) \left( \frac{f }{5 \times 10^{-4} \rm Hz} \right)^{2} \left( \frac{\tau}{3 \; \rm years} \right)^{1/2} \nonumber
\\ &\approx  70 \left( \frac{8 \; \rm kpc}{d} \right) \left( \frac{M_c}{28 M_\odot} \right)^{5/3} \left( \frac{f}{5 \times 10^{-4} \; \rm Hz 	} \right)^{8/3} \left( \frac{\tau}{3 \; \rm years} \right)^{1/2} \; .
\end{align}
For observations across the Milky Way with $d = 20$ kpc, we find,
\beq
SNR \approx 12 \times \left( \frac{20 \; \rm{kpc}}{d} \right) \left( \frac{\tau}{3 \; \rm years} \right)^{1/2} \; .
\eeq
Figure \ref{fig:LISA} shows the expected number of such Milky Way binaries as a function of their SNR.

\subsection{Confusion with cosmological sources}
A supermassive binary black hole at cosmological distances can possess similar values of strain amplitude and frequency to a Galactic binary black hole, thus masquerading as a Galactic source. However, this confusion can be eliminated by measuring the change in gravitational wave frequency as a function of time, $\dot{f}$. 

The time derivative of the gravitational wave frequency is given by \citep{ChirpMass}
\beq
\dot{f} =  \frac{96\pi^{8/3} f^{11/3} }{5} \left( \frac{ G}{c^3}   M_c \right)^{5/3} \propto M_c^{5/3} \; .
\eeq
Supermassive binary black holes possess chirp masses that are much greater than that of Galactic binaries. As such, their frequency changes at a much faster pace than Galactic binaries. 

\section{Electromagnetic flag}

\subsection{Binary black hole accretion in a hierarchical triple system} \label{subsection: Hierarchical triplet}

A binary black hole could accrete gas if it resides in a hierarchical triple system, where the third object is a main sequence star. The wind  of the third star would lead to accretion at the Bondi-Hoyle-Lyttleton rate \citep{BHL, BHL2},
\beq
\dot{M} = \frac{4 \pi G^2 M_{tot}^2 \rho_w}{\sqrt{ (c_w^2 + v_w^2)^3 } } \; ,
\eeq
where $M_{tot}=M_1 + M_2$, $\rho_w$ the mass density of the stellar wind, $v_w$ the wind speed, and $c_w \equiv (5k T_w/3 m_p)$ is the sound speed, with $T_w$ be the wind temperature. Scaling the wind parameters to solar values at a distance of $1$ AU and assuming an efficiency of $0.1$ for converting rest mass into radiation, the luminosity produced by the binary where both black holes are $\sim 30$ solar masses is,
\beq
L \approx 2 \times 10^{30} \left[ \frac{l}{1 \rm{AU}} \right]^{-2}  \rm erg \; s^{-1} \; ,
\eeq
where $l$ is the separation of the star from the binary. The maximum luminosity is given by the Eddington limit,
\beq
L_E \approx 10^{40} \left[ \frac{M_{tot}}{60 M_\odot} \right]   \rm erg \; s^{-1} \; .
\eeq
Due to the orbital motion of the black holes around the center of mass, the observed flux would be modulated by Doppler beaming. Assuming that the emitted flux, $F_{\nu0}$ scales with frequency as $F_{\nu0} \propto \nu^\beta$, the Doppler modulation is given by
\beq
F_\nu = D^{3-\beta} F_{\nu0} \;,
\eeq
where $F_\nu$ is the observed flux, and $D$ is the Doppler factor. To first order, the flux modulation is given by \citep{DanD},
\beq
\frac{\Delta F_\nu}{F_{\nu0}} \approx (3- \beta) \sqrt{\frac{G M_{tot}}{a}} \frac{ \cos{\phi} }{c} \sin{i} \; ,
\eeq
where $\phi$ is the orbital phase and $i$ the inclination.  

At the Eddington luminosity, the amplitude of the flux modulation of two $30M_\odot$ binary, is given by, 

\beq 
\Delta F \approx  0.6 \bigg[ \frac{10 R_\odot}{a} \bigg]^{1/2} \bigg[ \frac{\rm{pc}}{d} \bigg]^2 \rm{erg \; s^{-1} \; cm^{-2}} \; ,
\eeq
where $d$ is the distance to the object and we have assumed $\beta \sim 1$ \citep{spectralindex,DanD}. Given the flux sensitivity of XRM-Newton of $2 \times 10^{-15} \; \rm erg \; cm^{-2} \; s^{-1}$, for a semimajor axis of $a\sim10 R_\odot$, the flux modulation of these objects will be observable out to $d\sim 10 $ Mpc. 

Realistically, it is unlikely for such binaries to emit at the Eddington luminosity. For general Bondi-Hoyle-Lyttleton accretion, the flux modulation depends on the binary and stellar wind parameters,
\beq
\Delta F \approx \frac{0.1 \rho_w (3-\beta)c}{  d^2 \sqrt{ (c_w^2 + v_w^2)^3 }   } \sqrt{\frac{G^5 M_{tot}^5 }{a} } \sin i \; .
\eeq
Substituting the solar wind parameters at $d=1$ AU and the XRM-Newton sensitivity for $\Delta F$ yields for the observer distance of $d \sim 300$ pc at which a pair of $30 M_\odot$ black holes will be detectable. The X-ray surveyor\footnote{http://wwwastro.msfc.nasa.gov/xrs/}, a next generation x-ray observatory with a flux sensitivity of  $\sim 10^{-19} \; \rm erg \; cm^{-2} \; s^{-1}$, will be able to detect this flux modulation out to a distance of $\sim 30 $ kpc, which allows their detection throughout the entirety of the Milky Way galaxy.

Owing to the fact that most massive stars are in multiple systems \citep{Raghavan, Tokovinin}, and that there is precedent for X-ray binaries in triple systems \citep{triple1, triple2, triple3, triple4, triple5}, most binary black holes will likely possess a third companion. Further evidence of this comes from observations of superorbital modulations in high-mass X-ray binary systems, which could be caused by a third companion \citep{Superorbital1, Superorbital2}. However, only a fraction of binary black holes in a hierarchical triplet would host companions in the relevant mass range for accretion to be efficient. Since only companions with masses $\sim 1 M_\odot$ and above generates sufficient luminosity to be observed throughout the Milky Way, the percentage of observable systems is
\beq \label{eq:CompRatio}
f_{c} = \frac{\int_{1 M_\odot}^{100 M_\odot}     T_c(M) \Phi_M(M) dM   }{\int_{0.07 M_\odot}^{100 M_\odot} T_H \Phi_M(M) dM   } \; , 
\eeq
where $\Phi_M$ is the stellar IMF, $T_H \sim 1.4 \times 10^{10}$ yr is the Hubble time, and $T_c$ is the main sequence lifetime of the companion star, given by the broken power law \citep{startime1, startime2, startime3}
\beq
T_c = \tau m^{-\Gamma} \; ,
\eeq
where $(\tau, \Gamma) = (10^{10} \; \rm{years}, \; 2.5)$ for stars less massive than $3 M_\odot$ and $(\tau, \Gamma) = (7.6 \times 10^9 \; \rm{years}, \; 3.5)$ for more massive stars. The ratio in equation (\ref{eq:CompRatio}) takes into account both the stellar IMF and the fact that more massive stars live a shorter amount of time. Substituting the Chabrier IMF for $\Phi_M$, we obtain,
\beq
f_{c} \sim 6 \times 10^{-2} \; .
\eeq
This implies that out of the $\sim 300$ black holes with $a \le 10 R_\odot$ predicted by equation (\ref{eq:numberBHsimplest}), only a few systems would host the appropriate companion to be observable throughout the entirety of the Milky Way. This number is further diminished by the fact that only a fraction of all triple systems have the stars at a close enough distance, as the signal scales as $\rho_{w} \propto l^{-2}$. We therefore conclude that the most efficient method to detect these binaries is through their gravitational wave emission with LISA. 

\subsection{Tidal disruption flares from planets and asteroids}
Another source of electromagnetic activity could be associated with the tidal disruption of planets and asteroids by the black holes. White dwarfs and neutron stars are known to host rocky debris around them \citep{Andrew, WD2, WD3, WD4, PulsarPlanet, PulsarPlanet2, PulsarPlanet3}. Orbits around binary black holes can be chaotic and are subject to Kozai-Lidov oscillations, leading to an enhanced tidal disruption rate \citep{TDE1, TDE2, TDE3}.

For a planet or asteroid of mass $m_p$ and radius $r_p$ being tidally disrupted by a black hole of mass $M$, the length of the flare is defined as the time it takes for the emission to drop under the Eddington limit. This is given by \citep{TDEtime},
\begin{align}
t_{f} &\approx 1.9 \left( \frac{l_p}{l_t} \right)^{6/5}  \left( \frac{r_p}{R_\odot} \right)^{3/5} \left( \frac{m_p}{M_\odot} \right)^{1/5} \nonumber 
\\ &\;\;\;\;\;\;\;\;\;\; \times \left( \frac{\epsilon}{0.1} \right)^{3/5} \left( \frac{M_{tot}}{10^6 M_\odot} \right)^{-2/5} \; \rm{yrs} \; ,
\end{align}
where $l_p$ the pericenter distance and $l_t$ the tidal radius. Assuming a Neptune-like planet with $m_p \sim 10^{29}$ g and $r_p \sim 10^{29}$ cm, the flare time becomes $t_f \sim 1.4$ years. 

Since this timescale is much longer than the orbital timescale, lightcurves from these events will possess amplitude modulation due to the Doppler effect, as described in section \ref{subsection: Hierarchical triplet}. The Doppler modulation of such systems is bright enough to be detectable from throughout the Milky Way by existing telescopes such as the XMM-Newton and the Chandra X-ray Observatory. As the number of sources in the sky per unit time will depend on the duty cycle of such flares, a monitoring campaign of a large patch of the sky is required to find such flaring sources. Confusion with other sources would make the identification of such binaries difficult. 

\section{Conclusions}
By calibrating the population of binary black holes based on the merger rate infered by LIGO, we have found that LISA could detect a handful of such binaries in the Milky Way galaxy (see Figure \ref{fig:LISA}). A lack of detections will set constraints on the binary production mechanisms. 
  
We also considered electromagnetic flags of these tight binary black holes in the Milky way, and found them to have weaker observational prospects. Gravitational wave signals could be leveraged to provide both the system's masses and semimajor axis. 

\section{Acknowledgements}
The authors thank Josh Grindlay for comments on the manuscript. This work was supported in part by the Black Hole Initiative at Harvard University, funded by the John Templeton Foundation.

\section{Appendix}
\subsection{Power law source functions without a minimum or maximum scale}
Consider a nonzero, time-independent power-law source term $S(a) \propto a^n$ that extends all the way from $a=0$ to $\infty$. Unlike the example considered in the main text, we consider a case without a smallest injection scale or an upper cutoff scale. The binary black hole population with this source term obeys
\beq \label{eq:advecsource}
\frac{\dd \rho(a,t)}{\dd t} - \frac{\dd }{\dd a} \left[ K(M_1, M_2) \frac{\rho (a,t)}{a^3} \right] = K_2 a^n \; ,
\eeq
where $K_2$ and $m$ are constants that in principe can be either derived or estimated from observations. The general solution to this equation is given by 
\begin{equation}
\rho(a,t) = a^3 F\left[\frac{a^4}{4 K}+t\right] - \frac{K_2 a^{4+n}}{K(1+n)} \; ,
\end{equation}
where $F$ is an arbitrary function. As before, we search for a steady state solution by setting the function $F$ to be the constant $C$, giving 
\beq
\rho(a) = a^3 C- \frac{K_2 a^{4+n}}{K(1+n)} \; .
\eeq
In this case, the merger rate $R$ is equal to the flux in $a$-space at $a=a_m$ plus a term corresponding to the source term,
\beq
\rho(a_m) \frac{K}{a_m^3} + \sigma = R \; ,
\eeq
where $\sigma$ is the rate of binary black holes created per year with semimajor axis $a \le a_m$, given by
\beq
\sigma = \int_0^{a_m} K_2 a^m da = \frac{K_2}{n+1} a_m^{1+n} \; .
\eeq
In this case,
\beq
\rho(a_m) = \frac{a_m^3 R}{K} - \frac{K_2 a_m^{4+n}}{K(1+n)}  \; ,
\eeq
which allows us to deduce that $C=R/K$. Therefore, the number of Galactic binary black holes with semimajor axis $\le a$ is given by
\begin{align}
N(\le a) &= \frac{a^4 R}{4 K} - \frac{K_2 a^{5+n}}{K (1+n)(5+n)} 
\\ &= N_{\rm hom}(\le a) -  \frac{K_2 a^{5+n}}{K (1+n)(5+n)} \; ,
\end{align}
where $N_{\rm hom}(\le a)$ is the number of Galactic binary black holes in the source-less case. Note that when we set the source term to zero by using $K_2=0$, we will recover the sourceless solution. The effect of such a source term with $n > 0$, i.e. where there is a higher rate of binary black hole production at large semimajor axis, is paradoxically a suppression of the number of binary black holes in the steady state solution due to the requirement that the rate $R$ be kept unchanged. Since a source function that does not have a maximum injection scale is unphysical, the solution we obtain is also unphysical. In this case, there is a scale, $a_{\rm crit}$, above which $N(\le a)$ becomes negative. This calculation should be viewed only as a pedagogical toy model to illustrate an example of the ramifications of modifying one of our assumptions. 

\subsection{Population analysis for varying black hole masses: Power law source } 
We can repeat the analysis for the case of a power law source term. In this case we have,
\beq
\rho_K(a, K) = a^3 \frac{C}{K} - \frac{K_2 a^{4+n}}{K(1+n)} \; ,
\eeq
where, inspired by our previous solutions, we have explicitly written out the $1/K$ dependency of $C$ so that it is now a constant with respect to $K$. The merger rate is therefore given by, 
\begin{align}
R & = \int_{K_{min}}^{\infty} K_3 K^m   \left[ a_m^3 \frac{C}{K} - \frac{K_2 a_m^{4+n}}{K(1+n)} \right]  \frac{K}{a_m^3}   dK + \sigma\nonumber
\\ &= \sigma+\left[ C K_3 \frac{K^{m+1}}{(m+1)}  - \frac{K_3 K_2 a_m^{1+n} K^{m+1}}{  (1+n)(1+m) } \right]^{\infty}_{K_{min}} \; ,
\end{align}
which converges when $(m+1) < 0$ to 
\beq
R = \sigma + \left[ - C  \frac{K_3  K_{min}^{m+1}  }{(m+1) } + \frac{K_3 K_2 a_m^{1+n} K_{min}^{m+1} }{(1+n)(1+m) }  \right] \; . 
\eeq
Here $\sigma$ is again the rate of binary black holes created per year with semimajor axis $a \le a_m$. However, in order to be consistent with our choice of the black hole mass function, we need to take into account the fact that different amounts of binary black holes are created for different black hole masses. Equivalently, binary black holes with different $K$'s are produced at different abundances. As a result, $\sigma$ has to include an extra integral over $K$, 
\begin{align}
\sigma &= \int_{K_{min}}^{\infty} f_K \frac{K_2}{n+1} a_m^{1+n} dK \nonumber 
\\&=  \int_{K_{min}}^{\infty} K_3 K^m \frac{K_2}{n+1} a_m^{1+n}  dK \nonumber
\\&=  K_3 \frac{K_2}{(1+n)(1+m)} a_m^{1+n} \left[ K^{m+1}\right]^{\infty}_{K_{min}} \; . 
\end{align}
When $m+1<0$, this integral converges to,
\beq
\sigma = - \frac{K_3 K_2 a_m^{1+n} K_{min}^{m+1} }{(1+n)(1+m)} \; .
\eeq
Following through, this gives the number of binary black holes per unit semimajor axis to be,
\beq
\rho(a, K) = -\frac{R  (m+1)}{ K_3 K_{min}^{m+1} } \frac{a^3}{K}  - \frac{K_2 a^{4+n}}{K(1+n)} \; ,
\eeq
when $m<0$. Note that when we set the source term to zero through $K_2=0$, we will recover the sourceless solution. The number of binary black holes per unit semimajor axis is therefore,
\beq
\rho(a) =  \rho_{\rm{hom}}(a) + \frac{K_3 K_2 a^{4+n} K_{min}^m }{(1+n) m}  \;,
\eeq
where $\rho_{\rm{hom}} (a)$ is the solution in the sourceless case and the number of binary black holes with semimajor axis $\le a$ is given by, 
\beq
N(\le a) = N_{\rm{hom}} + \frac{K_3 K_2 a^{5+n} K_{min}^m }{(1+n) (5+n) m}  \; .
\eeq
where as before $N_{\rm{hom}}$ is the sourceless solution. This solution is again pathological due to the presence of a critical semimajor axis above which $N(\le a)$ becomes negative. 

\end{document}